# The "Bohr" Model for the High $T_c$ Superconductivity

Y. H. Kim[1*] and P. H. Hor[2**]

[1]Department of Physics, University of Cincinnati, Cincinnati, Ohio 45221-0011, U.S.A.

[2]Department of Physics and Texas Center for Superconductivity
University of Houston, Houston, Texas 77204-5005, U.S.A We propose a charge crystal model that captures all the essential physics of the high temperature superconductivity (HTS) in the long wavelength limit. Based on the recent transport and the far-infrared (far-IR) experiments, we argue that the three-dimensional (3D) ordering of the pinned two-dimensional (2D) square electronic lattice (EL) in each $CuO_2$ plane is the building block of HTS. Incorporating the physical picture derived from the neutron scattering experiments, we demonstrate that our model presents a coherent picture of the HTS. We suggest that the charge crystal model serves as a model for the microscopic theory and, hence, offers the key to the mechanism for the HTS.

* email address: kim@physics.uc.edu
** email address: phor@uh.edu

After two decades of extensive search for the underlying mechanism for the high superconducting transition temperature ($T_c$) of cuprate-based superconductors, the field of the high temperature superconductivity (HTS) has now become mature. During this time many key clues and experimental evidences that should be sufficient for building a microscopic model for HTS have been accumulated. However, in spite of the vast experimental data in hand, the microscopic theory of HTS puzzle remains unsolved. Granted this could be due to the complexity of the problem, but it would also be very likely that some key building blocks might have been overlooked as they were hidden behind various experiments on different high temperature superconducting materials and, therefore, quite subtle to be recognized. In order to solve this puzzle, we need to "re-search" all the pieces in hand and put together the key pieces to bring about a coherent picture that captures the essential physics of the HTS.

Our starting place, for the charge channel, is the "superconducting dome" that exists in a narrow range of 2D hole density (p) [1] in the generic p versus temperature (T) electronic phase diagram [2]. For cation (Sr) doped $La_{2-x}Sr_xCuO_4$ (SD-La214) the superconducting dome begins at $p_c \approx 0.06$ and ends at $p_c \sim 0.25$, reaching the maximum $T_c$, typically less than 40 K, at $p \approx 0.16$. However, type of the dopants affects this general behavior greatly; it was found in the metastable, pure anion (O) doped $La_2CuO_{4+\delta}$ (OD-La214) that the $T_c$ can reach 45 K [3, 4], and even ~ 60 K in an inhomogeneous thin film [5]. Furthermore, for the spin channel, one must also inquire into the specific role of the ubiquitous antiferromagnetic (AF) Cu atom exchange interaction in the cuprate physics in general and the connection between the HTS and the static spin density wave (SDW) order development at $T_{SDW} = T_c$ specifically [6 - 9].

The prospect of the solution comes from recognizing the existence of some "magic" doping numbers $p = m/n^2$ with nonzero integers m and n where $m < n^2$ and the two-component nature of doping induced holes (coexistence of the itinerant and the localized holes) in various cuprates. Note that there is a simple geometrical meaning of the magic doping concentration, namely, on the Cu-Cu lattice basis they are the exact hole concentration for charge ordered arrangement of 2D square lattices in the $CuO_2$ planes, the 2D electronic lattices (EL's). There are many scattered early indications of "magic" doping concentrations, localization of doped holes, and non-equivalent Cu-sites due to charge inhomogeneity in the under-doped cuprates [10].

The first thermodynamic evidence for the EL formation and the two-component nature of the doping-induced holes came from the electrochemical doping studies of SD-La214 and cation/anion (Sr/O) co-doped $La_{2-x}Sr_xCuO_{4+\delta}$ (CD-La214). A change of doping efficiency, from 2 holes per doped oxygen ion for $p < p_c$ into 1.3 holes per doped oxygen ion for $p > p_c$, that signals a sudden increase in the chemical potential at a critical hole concentration $p_c \sim 0.06$ was found [11]. A sudden increase of the chemical potential indicates either a dramatic change of global electronic structure in which all doped holes are involved or a formation of a "protected" electronic state at $p_c$ such that further doping will force the holes to occupy a different electronic state. Experimentally we observe, for $p < p_c$, the hole density depends linearly on the excess oxygen content ($\delta$) as $p = 2\delta$ which extrapolates to zero. However $p = 1.3\delta + 0.019$ for $p > p_c$. Since $p_c$ is independent of the nature (Sr or O) of the dopants, it must be a genuine intrinsic, globally uniform property of the doped holes in the $CuO_2$ plane. Therefore, this finite intercept of $p = 0.019$ at $\delta = 0$ indicates that the excess doped-holes ($\Delta p = p-p_c$) have to occupy a different electronic state that is built on the underlying "protected" electronic state. At this $p_c$ the metal-to-insulator transition occurs in the $CuO_2$ at high temperature and the system eventually becomes superconducting at a lower temperature for $p > p_c$.

These observations suggest that there exists a distinct electronic state at $p_c \sim 0.06$ and HTS is built on this robust electronic state by further doping. Therefore, we, due to the vicinity of $p_c \sim 1/16$ to the magic doping concentration and some early experimental suggestions of 2D square lattice formation [10], conjectured that the $p_c \sim 0.06$ electronic state to be a $4a$ x $4a$ (where *a* is the in-plane Cu-Cu distance) square Wigner lattice ($p_c = 1/4^2 = 0.0625$) with excess $\Delta p$ holes moving on top of this underlying 2D EL [12]. In this way, only fraction of the doping-induced holes will enter the condensate state. This two-component nature of the doping-induced holes was further supported by the far-infrared (far-IR) studies of the optimally doped SD-La214 [13] and Bi2212 [14] in which it was found that only $\sim 20\%$ of the total spectral weight was in the condensate state. Furthermore the most recent observations of a series of electronic phase separations due to two hierarchical series of magic doping concentrations like $p = (1/4)^2, (1/3)^2, 2(1/4)^2$, $3(1/4)^2$, $2(1/3)^2$, $4(1/4)^2$... in both normal and superconducting states have clearly demonstrated the fundamental role of the 2D EL in the cuprate physics [15]. There are also many recent compelling experimental evidences that exhibit magic doping that is consistent with 2D EL formation. For instance, scanning tunneling microscopy (STM) studies began to unravel the evidences for the 2D electronic ordered states in $Bi_2Sr_2CaCu_2O_{8+d}$ [16]. Furthermore the in-plane resistivity studies of SD-La214 revealed rational doping fractions consistent with checkerboard-type ordering of Cooper pairs [17].

Studies of the phase diagram of well-annealed equilibrium samples of OD-La214, where the dopants are mobile for T > 200 K, revealed that there are only two $T_c$'s, one at ~ 15K and the other at ~ 30K, in the entire under-doped regime [18]. Further studies of the electronic phase diagram and pressure dependences of CD-La214 showed that these were two distinct (one at $T_{c1}$ = 15 K and the other at $T_{c2}$ = 30 K) intrinsic "electronic" superconducting phases [19]. Focusing on the two intrinsic $T_c$'s, the connection between the EL order and $T_{c1}$ and $T_{c2}$ was established in the far-IR charge dynamics studies of the CD-La214 [20, 21] by the observation and clear association of two collective modes $\omega_{G1}$ ~ 23 cm$^{-1}$ and $\omega_{G2}$ ~ 46 cm$^{-1}$ to $T_{c1}$ and $T_{c2}$, respectively. All the above experiments were done on the polycrystalline samples with the oxygen dopant always present. The same consistent physical picture was confirmed in the far-IR [22] and magnetic [23] studies of high quality SD-La214 single crystals.

In the metastable OD-La214 [4], where the oxygen dopants were excessively injected electrochemically, it was found that there present the collective modes at $\omega_{G1}$ ~ 23 cm$^{-1}$, $\omega_{G2}$ ~ 46 cm$^{-1}$, and $\omega_{G3}$ ~ 72 cm$^{-1}$ corresponding to $T_{c1}$ = 15 K, $T_{c2}$ = 30 K, and $T_{c3}$ = 45 K superconducting phases respectively. Since a collective mode is an unmistakable fingerprint of the presence of an EL [24], the observation of the collective mode (Goldstone mode) at a finite frequency is significant. If the EL were not pinned, the $\omega_G$ would appear at $\omega = 0$ and entire EL would slide under the influence of an electric field. If the EL is pinned, then there exists a gap in the Goldstone mode and we will observe $\omega_G$ at a finite frequency and the EL becomes an insulator. Therefore, $\omega_G^2$ measures the strength of the commensuration pinning of the EL. Our experimental observation indicates that $\omega_G$ in SD-La 214 increases discretely in steps of ~ 23 cm$^{-1}$ as the hierarchical order of the EL increases from $p_{c1}$ = 1/16 EL to the next ones, suggesting that $\omega_G$ is intimately tied to both the EL and the superconductivity.

The sum rule analysis of the far-infrared conductivity of the CD-L214 suggests that only extremely small fraction (< 1%) of the doping-induced holes are responsible for the normal state charge transport in the $CuO_2$ plane (ab-plane) with a surprisingly small scattering rate $\Gamma_{ab}$ ~ 10 cm$^{-1}$ [20 - 22] and almost all the doping-induced holes (> 99%) contribute to the collective modes and the carrier density involved in the $\omega_{G2}$ mode is twice higher than that of the $\omega_{G1}$ mode. Subsequently, the corresponding symmetry of the EL of the $\omega_{G1}$ mode was identified as p(4x4) ($p_{c1}$ = 1/16 = 0.0625) and that of the EL of $\omega_{G2}$ as c(2x2) ($p_{c2}$ = 2/16 = 0.125) symmetry (see Figure 1(a) and 1(b)). Then, the $T_c$ = 45 K superconducting phase accompanying $\omega_{G3}$ mode in the metastable $La_2CuO_{4+\delta}$ must have the $p_{c3}$ = 3/16 EL symmetry as depicted in Figure 1(c). Then, all the intermediate $T_c$'s in the superconducting dome are resulted from the mixture of these three basic EL's and one may calculate their composition by applying a lever rule [25]. For instance, in p = 0.07 CD-La214 ($T_c$ ~ 28 K), 13.3 % of the $p_{c1}$ EL coexists with the $p_{c2}$ EL [21]. The optimally doped SD-La214 (p ~ 0.16 and $T_c$ ~ 38 K) consists of 46.7% of the $p_{c2}$ EL and 53.3% of the $p_{c3}$ EL. It is also noteworthy that the linshape of $\omega_G$'s of the single crystalline SD-La214 are much sharper than those of the electronically homogeneous CD-La214, but appear at the lower frequencies [20]. This indicates that while the macroscopic domains of individual EL's are more easily formed in the polycrystalline samples, the single crystalline SD-La214 has interwoven patches of strongly coupled EL's. Therefore the pinning is obviously derived from the intrinsic properties of the 2D $CuO_2$ planes.

The presence of the EL dictates the free carrier transport in a unique way. Since the majority of the doping-induced holes are condensed into the insulating EL at p = $p_c$, the number of the excess holes, $\Delta p$ available for the charge transport is small. These $\Delta p$ holes naturally occupy the energetically minimum interstitial sites of the EL and the zero-point energy of the carriers in the interstitial sites becomes broadened into a band (the Coulomb band) since the free carriers in the interstitial sites can "see" the energetically equivalent sites in the directions of the neighboring energetically equivalent sites [20, 21]. The free carriers in the coulomb band will not experience the scattering with the phonons of the underlying $CuO_2$ lattice, hence, nearly dissipation free.  A similar interstitial band idea was considered first by de Wette in the study of the melting of 3D Wigner lattice [26].  Therefore, the ab-plane scattering rate $\Gamma_{ab}$ should be doping-independent and show the linear T-dependence in the ab-plane resistivity ($\rho_{ab}$) for T down to T ~ $0.2\theta_D$, where $\theta_D$ is the Debye temperature of the EL, because of the dominating scattering with the acoustic phonons of the EL.  For example, $\rho_{ab}$ would be linear down to T ~ 12 K for the p(4x4) EL [20].

Another interesting consequence of this EL ground state is that we anticipate the metallic charge transport at moderate T even at p ~ 0.01 as long as the extra carriers occupy the interstitial sites of the p(10x10) EL ($p_c$ = 0.01) as observed in the ab-plane transport study of SD-La214 single crystal (p ~ 0.01) [27].  As p increases, the development of a series of 2D EL is expected whenever p becomes commensurate with the $p_c$ of the next higher order EL and the metallic charge transport results when $\Delta p \neq 0$. This is the physical origin of the magic number observed in various experiments as a function of p [10, 11, 16, 17, 23].  All the square EL formed p < $p_{c1}$ = 1/16 = 0.0625 become insulating as $T \to 0$ when the excess carriers can afford to pay for the Coulomb energy by occupying the lattice sites of the higher order EL symmetry.  Then the true insulator-metal transition should occur at $p_c$ ~ 0.06 where the p(4x4) lattice ($p_{c1}$ = 0.0625) can support the excess free carriers $\Delta p$ = p – 0.0625 as $T \to 0$ and the system begins to support the superconductivity [21]. When $p \to 0.12$, all the holes will condense into the $p_{c2}$ EL in principle, depriving all the free carriers of the $p_{c1}$ = 1/16 EL.  The dip at p ~ 0.12 in the superconducting dome is the reminiscence of the $p_{c2}$ = 2/16 EL formation.  As p increases beyond p ~ 0.12, $p_{c3}$ EL begins to emerge as some of the free carriers at the interstitial sites condense to form a new lattice and coexists with the $p_{c2}$ EL until p reaches $p_{c3}$.  As p increases further, the p (2x2) EL ($p_{c4}$ = 1/4) would begin to condense and for p > $p_{c4}$ the EL is unstable against Fermi-liquid ground state and melting of the EL should occur in SD-La214 and the superconductivity ceases to exist.

In this EL picture, the so-called far-IR pseudogap seen at $\omega$ ~ 400 $cm^{-1}$ at all T in the under-doped regime [28] is the single particle excitation gap of the 2D EL.  As the doping level increases, this single particle excitation gap becomes smeared by the free carrier contributions.  Also, the doping-independent energy gap measured at ~ 0.05 eV (400 cm$^{-1}$) in the ARPES experiment [29, 30] is the single particle excitation gap of the 2D EL. The manifestation of the pseudogap in the transport at ~ 200 K comes from the completion of the development of the 2D EL at ~ 200 K as seen in Sr/O co-doped SD-La214 experiment [20].

After sorting out the essential issues of the normal state properties of the cuprates within the 2D EL model, the next important point to address is the implication of the 2D EL for the c-axis charge transport because it has been generally believed that the normal state c-axis charge transport is incoherent: A critical experimental justification for all the non-Fermi liquid approaches to HTS. The remarkable recent far-IR observation [31] of the nearly dissipationless normal state charge transport along the c-axis with the c-axis scattering rate $\Gamma_c$ ~ 13 cm$^{-1}$, comparable to the ab-plane scattering rate $\Gamma_{ab}$ ~ 10 cm$^{-1}$, in p = 0.07 SD-La214 single crystal clearly suggests that the c-axis charge transport is intrinsically coherent and shares the same physical origin as the ab-plane charge transport in the 2D EL ground state. This leads us to a conclusion that the 2D EL's in the CuO$_2$ planes must undergo, although incomplete, a 3D ordering. Furthermore, in order to develop the same type zero-point energy band along the c-axis as in the ab-plane, the EL's on the neighboring planes must line up with each other. Therefore, we concluded that while HTS is based on the 3D ordered EL the essential cuprate physics has to be dominated by the 2D EL in each CuO$_2$ plane.

Since the same free carriers are responsible for the charge dynamics both in the ab-plane and along the c-axis and the unscreened ab-plane free carrier plasma frequency ($\omega_p^{ab}$) is much higher than the unscreened c-axis free carrier plasma frequency ($\omega_p^c$) and yet the normal state plasma edge appears at the same frequency for both the ab-plane and the c-axis reflectivities of p = 0.07 sample, the dynamic masses must be highly anisotropic and the screening must also be highly anisotropic. From $\omega_p^{ab}$ ~ 511 cm$^{-1}$ and $\omega_p^c$ ~ 78 cm$^{-1}$ for p = 0.07 and $\omega_p^{ab}$ ~ 542 cm$^{-1}$ and $\omega_p^c$ ~ 76 cm$^{-1}$ for p = 0.09 at T = 300 K [20, 31], the ratio between the c-axis dynamic mass (m$_c$) and the ab-plane dynamic mass (m$_{ab}$), m$_c$/m$_{ab}$ ~ 43 for p = 0.07 and ~ 51 for p = 0.09. Then, the intrinsic transport anisotropy ratio between the ab-plane resistivity ($\rho_{ab}$) and the c-axis resistivity ($\rho_c$) should be $\rho_c/\rho_{ab}$ ~ 56 for p = 0.07 and ~ 123 for p = 0.09 sample at T = 300 K. We anticipate that this intrinsic coherent transport anisotropy would continue to grow as the doping increases because $\Gamma_c$ increases with doping [31] but the ab-plane scattering rate $\Gamma_{ab}$ remains more or less the same [20]. Therefore, the normal state of high $T_c$ cuprate is a peculiar highly anisotropic 3D metal built on the charge crystal due to 3D ordering of 2D EL.

Next we wish to address the issues concerning the superconducting state within the 3D charge crystal model. In order to achieve the superconducting state, we need to answer the following two fundamental questions:

1. *What is the "glue" for the superconducting pairs?* – One novel feature of the 2D square EL ground state is that the free carriers that are "riding" on the 2D EL naturally form spin-singlet Cooper pairs because they experience the negative dielectric screening provided by the EL in the frequency region between the Goldstone mode ($\omega_G$) and the plasma frequency of the EL ($\Omega_{EL}$) [20]. This dielectric pairing is non-retarded, real-space pairing. Because of this negative screening, the Coulomb interaction between two electrons becomes attractive and, therefore, for $T < T_o \sim \Omega_{EL} \sim 250$ K, the free carriers in the EL form spin singlet Cooper pairs. It is interesting to note that such a dielectric pairing possibility was suggested first by Bagchi [32]. These pairs will enter into the superconducting condensate when the phase coherence of the Cooper pairs develops at $T = T_c$. It should be noted that since we observe no Goldstone modes along the c-axis of charge crystal, this pairing is strictly 2D confined in the ab-plane. Therefore, the phase coherence transition of the Cooper pairs into the superconducting state at $T = T_c$ requires a mechanism of a different physical origin.

2. *How does the system turn on its bulk superconductivity?* – The neutron scattering experiments have established that the development of the static incommensurate SDW order [6 - 9] is directly connected to the HTS. Although it was originally alluded to the spin-mediated pairing, it is important to note that the charge ordering and, hence, the Cooper pairing occurs at a much higher temperature than the magnetic ordering. Since the free carriers in the 3D charge crystal naturally form Cooper pairs when the long-range order of the corresponding EL fully develops at $T \sim 200$ K [20], all it requires for forming the superconducting condensate is the phase coherence development among the spin-singlet Cooper pairs. We propose that the spin-singlet pairs of the preformed Cooper pairs acquire the phase coherence when the spins undergo the SDW transition at T = $T_{SDW}$, and the superconductivity follows.

The presence of the EL modifies the AF spin environment of the $CuO_2$ plane in a special way. After the Heisenberg AF ground state in the $CuO_2$ plane is completely destroyed by the hole doping at p ~ 0.02 [33], the system enters into a highly unusual spin glass state [34] until the more stable charge-ordered ground state is formed at $p_{c1}$ corresponding to p(4x4) EL. During the course of the spin dynamics changes with doping, the hierarchical formation of a series of square EL's is involved with increasing p as discussed before. Now as the new 2D EL ground state of lattice constant L is established on the $CuO_2$ plane, there develops at the same time a 2D array of AF spin blocks with an effective spin, $S^{block}$ = 1/2 as shown in Figure 2. Note that the effective spin for p(3x3) block is zero. In this arrangement the inter-block AF coupling $J_{block}$ becomes renormalized as $J/N_s$ where J is the nearest neighbor super-exchange integral in the $CuO_2$ plane and $N_s$ is the number of spins contained in each spin block. Then, when a free hole with spin s = 1/2 is placed on the EL, the spin of the hole $\vec{s}$ at position $\vec{r}$ antiferromagnetically couples to the underneath block spin $\vec{S}_i^{block}$ at $\vec{R}_i$ as $K \vec{s} \cdot \vec{S}_i^{block} \delta(\vec{r} - \vec{R}_i)$ (K > 0). Therefore, when the block spins undergo the SDW transition at T = $T_{SDW}$, the preformed Cooper pairs in the Coulomb band acquire the phase coherence of the SDW and the superconductivity results. In this model the Cooper pairs formed in the charge crystal based on p(3x3) EL cannot acquire the phase coherence, hence, does not support the superconductivity.

Since the AF spin dynamics of the doped cuprates is inherited from that of the undoped parent cuprates and the stacking arrangement of the magnetically ordered planes and the spin direction is identical to the pristine $La_2CuO_4$ [7, 35], one may write down a simple effective block spin Hamiltonian with the nearest neighbor block spin interaction $J_{block}$ as

$$H_{spin} = J_{block} \left[ \sum_{i,j} \vec{S}_i^{block} \cdot \vec{S}_j^{block} + \lambda_\perp \sum_{i,j'} \vec{S}_i^{block} \cdot \vec{S}_{j'}^{block} \right].$$

Here $\lambda_\perp$ is the coupling between the block spin at $i$ in the $CuO_2$ plane and the block spin at $j'$ of the two neighboring $CuO_2$ planes. This inter-plane spin coupling is essential to drive the magnetic ordering transition no matter how small it may be. The SDW ordering temperature $T_{SDW}$ of $H_{spin}$ can be found from the work done by Soukoulis *et al.* [36] as $k_B T_{SDW} = (4\pi/3) J_{block} / \ln(32/\lambda_\perp)$ in the $\lambda_\perp \to 0$ limit. Since the superconducting phase transition occurs at $T = T_c = T_{SDW}$ and $N_s$ can be written as $N_s \approx (L/a)^2$ ($a$ = Cu-Cu distance) in 2D (see Figure 2), $J_{block}$ may be written as $J_{block} = J/(L/a)^2$ to find $T_c$ for a specific EL crystal ground state at $p = p_c$ as $k_B T_c = (4\pi/3) p_c J / \ln(32/\lambda_\perp)$ using $p_c = (L/a)^{-2}$. Notice that the $T_c$ is linearly proportional to $p_c$ as observed in the far-IR experiment [20] and that $\lambda_\perp$, the residual inter-plane magnetic coupling, plays an important role to induce the superconducting transition. This is consistent with the importance of "incomplete suppression" of AF order to cuprate physics suggested in Ref. [37] and with the importance of weak interlayer coupling to spin-glass phase [38].

When we use the same $\lambda_\perp$ of the undoped La$_2$CuO$_4$ [39], $\lambda_\perp \approx 3 \times 10^{-5}$ and $J \approx 130$ meV, for SD-La214, we overestimate the $T_c$'s by a factor of two as $T_c \sim 29$ K for the $p_{c1}$ EL crystal. However, since the mobile holes are responsible for the dramatic reduction of $\lambda_\perp$ for p up to 0.02 [40], it is conceivable that $\lambda_\perp$ may substantially further decrease with the doping. For example, if we choose $\lambda_\perp = 1.2 \times 10^{-10}$ for SD-La214, we have the correct $T_c = 15$ K for the $p_{c1} = 1/16$ EL crystal. Then the $p_{c2} = 2/16$ EL crystal would support $T_c = 30$ K and the $p_{c3} = 3/16$ EL crystal $T_c = 45$ K as experimentally observed. Likewise, if we consider the reduction in $\lambda_\perp$ from 0.079 [41] to $\lambda_\perp \sim 8 \times 10^{-5}$ with doping for YBCO, we will have another $T_c$ sequence with $J = 120$ meV. Namely $T_c = 30$ K for the $p_{c1} = 1/16$ EL, $T_c = 60$ K for the $p_{c2} = 2/16$ EL, and $T_c = 90$ K for the $p_{c3} = 3/16$ EL in YBCO, indicating that, based on our charge crystal model, the 60K and 90K phases are purely electronic. This agreement with the experiment is clearly satisfactory. Nevertheless, it is necessary to carry out a rigorous calculation of $T_{SDW}$ in the EL ground state and the spin-pair condensation energy of the Cooper pairs using the full Hamiltonian including both AF spins in the presence of EL and the interaction term $K \vec{s} \cdot \vec{S}_i^{block} \delta(\vec{r} - \vec{R}_i)$ while maintaining the spin singlet state of each Cooper pair.

When the SDW ordering sets in, the neutron scattering experiment should be able to measure the spin excitation energy gap of the SDW arising from the anisotropy.

However, the SDW gap obtained by diagonalizing $H_{spin}$, $\Delta_{spin} = 2 p_c J \sqrt{\lambda_\perp}$ in $q \rightarrow 0$ limit [42] would be too small to be seen in the inelastic neutron scattering experiment. In our model, due to the small number of the free carrier density in the underdoped regime, the spin pair condensation energy, $\delta E_{spin}$, measurement requires a sensitive probe. For SD-La214 the spin pair condensation energy gap will not be easily seen in the inelastic neutron scattering experiment until the doping reaches the optimal level where the $T_c$ is maximum. This is the reason why the spin pair condensation energy gap is observed only for the optimally doped or nearly optimally doped SD-La214 in the inelastic neutron scattering experiment [43]. For SD-La214 ($T_c \sim 38.5$ K), the spin pair condensation energy gap was found to be momentum-independent (isotropic) and its size is $\delta E_{spin} = 6.7$ meV [43].

One can also verify the validity of the EL based spin model through the measurement of the incommensurability $\eta$ via the elastic neutron scattering experiments and the SDW collective mode through the inelastic neutron scattering measurements. Without mixing of the two EL's, the elastic neutron scattering should find $\eta \sim 0.06$ for the $T_c = 15$ K phase and $\eta \sim 0.12$ for the $T_c = 30$ K phase. We suggest that the topology of the SDW is grid-like rather than stripe-like and the orientation of the SDW grid of $T_{SDW} = 30$ K is rotated by 45° from that of the SDW grid of $T_{SDW} = 15$ K by reason of the p(4x4) and c(2x2) EL symmetries. For the $p_{c3} = 3/16$ EL, we could have either diagonal spin stripes within the c(2x2) charge stripes of spacing shown in Figure 1(d) or grid matching the EL diagram depicted in Figure 1(c). However, the elastic neutron scattering studies of stage-4 $La_2CuO_{4+\delta}$ ($T_c \sim 42$ K) [7], in which the $p_{c3} = 3/16$ EL makes up 80% of its EL, indicate that the SDW direction is tilted only by 3° from the Cu-O-Cu direction, which cannot come from the diagonal spin stripes resulting from the diagonal charge arrangement depicted in Figure 1(d). Moreover, their inter-plane SDW ordering study indicates that the SDW's on neighboring planes must be antiferromagnetically arranged and parallel with the EL [7]. The source of the inter-plane spin ordering via virtually vanishing $\lambda_\perp$ comes from the 3D ordering of the EL because the SDW texture is closely related to the structure of EL's. Therefore, the 2D grid geometry of SDW would be consistent with the elastic neutron scattering data. The 3° tilting may come from the contribution of 20% of the $p_{c2}$ = 2/16 EL. The same will go for YBCO system as well. Such a SDW grid possibility was suggested first by Lee *et al.* [7].

Since the block-spin configuration is interlocked with the EL (see Figure 2), there should exist the SDW collective mode tied to the Goldstone mode of the EL. For instance, the ~ 3 meV (24 cm$^{-1}$) resonance peak [44] and the 9 meV (72 cm$^{-1}$) resonance peak [45] that develop at T = $T_c$ in the inelastic neutron scattering measurement of SD-La214 system are the SDW collective modes associated with the corresponding Goldstone mode of the $p_{c1}$ = 1/16 EL and $p_{c3}$ = 3/16 EL respectively. By the same token, the famous 41 meV (330 cm$^{-1}$) neutron resonance peak in $T_c$ = 90 K YBCO [46] and the ~ 30 meV resonance observed in $T_c$ = 63 K YBCO [47] can be identified as the SDW collective modes associated with the Goldstone modes of the $p_{c3}$ = 3/16 EL and $p_{c2}$ = 2/16 EL in YBCO respectively [48]. Therefore, we predict that there should also exist a Goldstone mode associated with the $p_{c1}$ = 1/16 EL in YBCO at ω ~ 110 cm$^{-1}$ and the corresponding neutron resonance peak that develops at ~ 14 meV at T = $T_c$ in underdoped YBCO, especially for $T_c$ ~ 30 K sample. As is the case in SD-La214, depending on the doping level, we anticipate coexistence of a series of Goldstone modes with varying relative oscillator strengths in YBCO due to the electronic disorder inherent to the chemically doped cuprate single crystals. [48, 49]

In the development of the hierarchical order of the EL in the CuO$_2$ plane, there is no obvious reason to have the square EL's with the long-range Coulomb interaction alone. In fact, a triangular EL is energetically favored [50]. When we introduce the strong electron-lattice interaction, the symmetry of an EL may be dictated by the symmetry of the underlying CuO$_2$ lattice. However, with the Coulomb interaction and the electron-lattice coupling alone, nothing prevents from forming another class of square EL order such as p(3x3) ($p_c$ = 1/9 = 0.11) symmetry. It was found that this p(3x3) lattice is unstable against the c(2x2) lattice at low T in SD-La214 [3, 4]. We believe that the AF spin dynamics plays a crucial role in stabilizing the square EL with specific symmetry For example, YBCO-like systems and other single layer systems (Hg1201 and Bi2212) are able to reach $T_c$ ~ 90 K at $p_c$ ~ 0.19 by stabilizing the p(2x2) EL ($p_{c4}$ = 4/16 = 0.25) [1], which was not possible for SD-La214. We believe that the spin ordering plays a crucial role in stabilizing the p(2x2) EL and, therefore, the interlayer spin coupling plays a crucial role in reaching the higher $T_c$ through the stabilization of the higher order EL.

Finally, in our model, since the c-axis Goldstone mode frequency is higher than the c-axis single particle excitation energy gap by the fact that the inter-EL distance between the neighboring $CuO_2$ planes is rigidly fixed by the inter-plane spacing, the c-axis dielectric function of the 3D ordered EL is always positive for frequencies below the single particle excitation gap and, hence, the dielectric screened pairing channel is absent along the c-axis. As demonstrated through the far-IR experiment [31], even though the 3D ordering of the EL develops and enables the nearly dissipationless normal state c-axis transport, the superfluid fraction along the c-axis is always partial even at the optimal doping because of the absence of the Cooper pairing interaction in the c-axis charge transport channel. We believe this partial superfluid fraction along the c-axis is an experimental manifestation of the missing pairing glue to form Cooper pairs along the c-axis.

In summary, purely based on the experimental facts, we have constructed a novel charge crystal model for HTS where very small fraction of free holes moving in a 3D hole crystal. We have demonstrated that our model presents a coherent picture of HTS in both normal and superconducting states. We suggest that our model can serve as a working phenomenological charge model for the further development of the microscopic theory of HTS.

We thank Eugene Kim for drawing Figures 1 and 2. P. H. H. is supported by the State of Texas through the Texas Center for Superconductivity at University of Houston.

**Figure Captions**

Figure 1. Schematic diagrams of the two-dimensional square electronic lattice (EL) in the $CuO_2$ plane at various doping levels. (a) p(4x4) EL ($p_{c1}$ = 1/16), (b) c(2x2) EL ($p_{c2}$ = 2/16), (c) $p_{c3}$ = 3/16 square EL, and (d) possible one-dimensional stripe arrangement at $p_c$ = 3/16. The smaller dots indicate the copper sites. The larger dots represent the sites occupied by the holes.

Figure 2. The block spin arrangements (drawn by the dashed lines) of the antiferromagnetically ordered spins (indicated by the arrows) in the presence of the EL. Notice that the block spins would be ferromagnetically arranged for $T > T_{SDW}$, which should give rise to a net local magnetic moment for $T > T_c$ even though the system is globally antiferromagnetic because the spins on neighboring planes are antiferromagnetically arranged. (a) The block spin arrangement with the p(4x4) EL. (b) The block spin arrangement in the c(2x2) EL environment. (c) The block spin arrangement with the p(3x3) EL. The open circles indicate the sites occupied by the holes.

**(a) $p_c = 1/16 = 0.0625$**

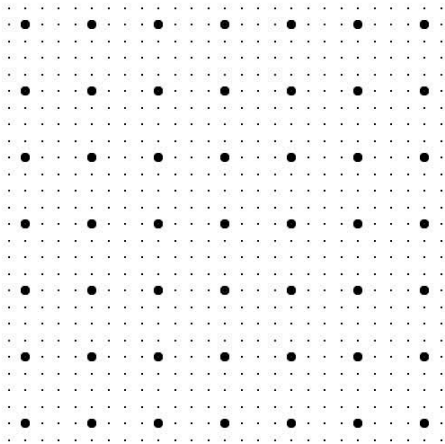

**(b) $p_c = 2/16 = 0.125$**

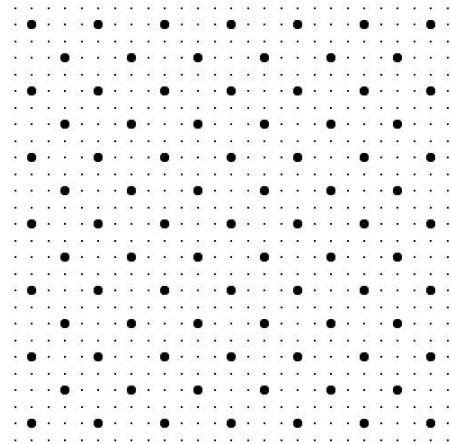

**(c) $p_c = 3/16 = 0.1875$**

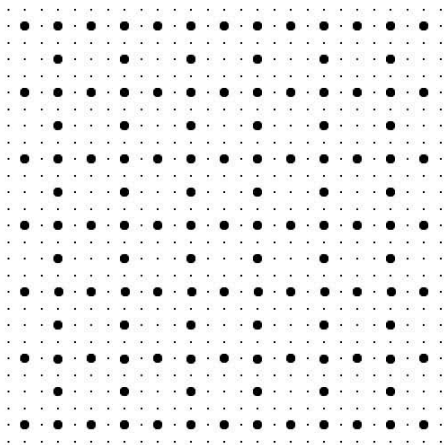

**(d) $p_c = 3/16 = 0.1875$**

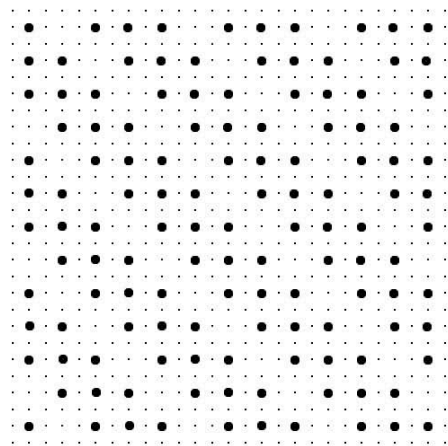

FIGURE 1

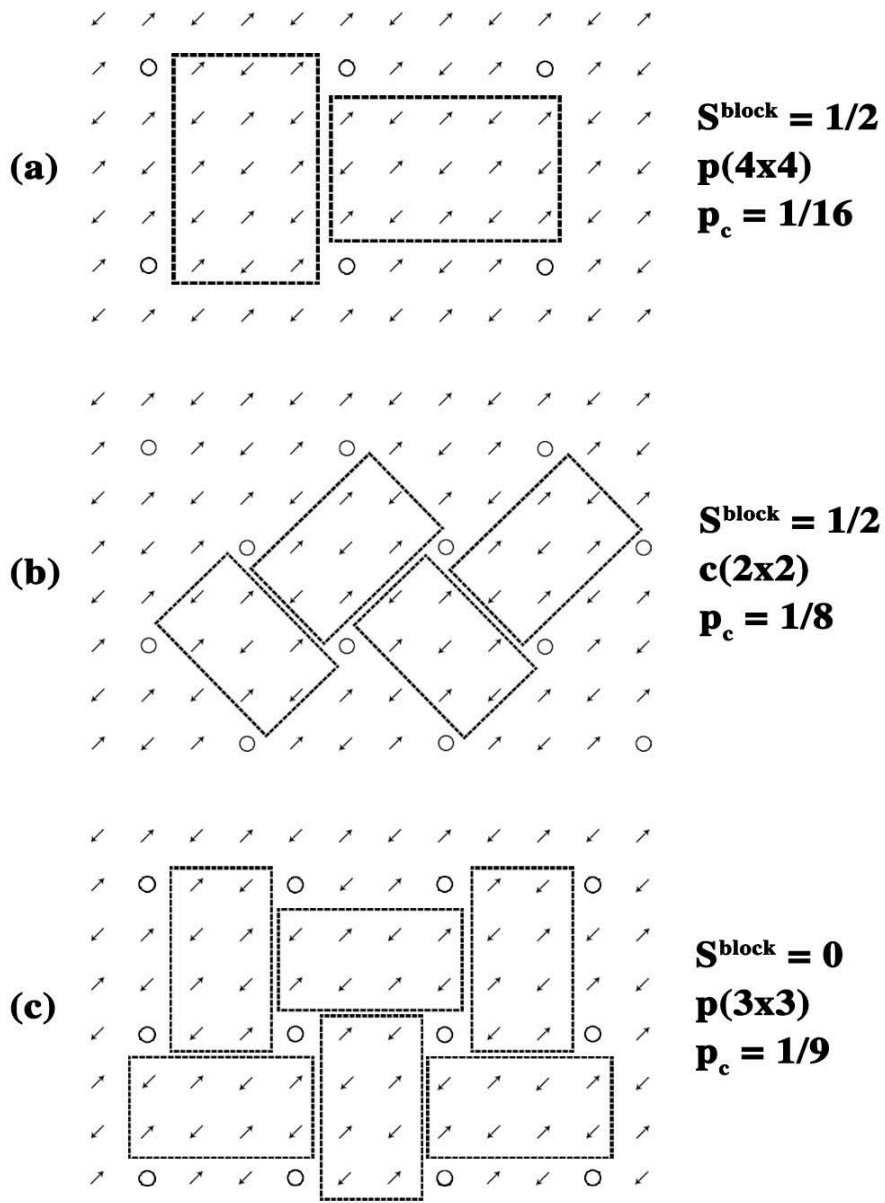

FIGURE 2